\pdfoutput=1

\documentclass[11pt]{article}

\usepackage[final]{acl}

\usepackage{times}
\usepackage{latexsym}
\usepackage{setspace}
\usepackage[T1]{fontenc}

\usepackage[utf8]{inputenc}

\usepackage{microtype}

\usepackage{inconsolata}

\usepackage{graphicx}
\usepackage{inconsolata}
\usepackage{graphicx}
\usepackage{hyperref}
\usepackage{multirow}
\usepackage{float}
\usepackage{booktabs}
\usepackage{makecell}
\usepackage{algorithm}
\usepackage{algorithmicx}
\usepackage{algpseudocode} 
\usepackage{stfloats}
\usepackage{amsmath}  
\usepackage{xcolor}
\usepackage{colortbl}

\usepackage{url}
\usepackage{xurl}
\usepackage{tabularx}
\usepackage{enumitem}
\usepackage{amssymb}
\usepackage{pifont}

\title{Automate Strategy Finding with LLM in Quant Investment}

\author{
Zhizhuo Kou\textsuperscript{1}, 
Holam Yu\textsuperscript{2}, 
Junyu Luo\textsuperscript{3}, 
Jingshu Peng\textsuperscript{1}, 
Xujia Li\textsuperscript{1},\\
\textbf{Chengzhong Liu}\textsuperscript{1},
\textbf{Juntao Dai}\textsuperscript{3},
\textbf{Lei Chen}\textsuperscript{1, 2},
\textbf{Sirui Han}\textsuperscript{1}\thanks{Corresponding author.} \thanks{Project leader}
\textbf{Yike Guo}\textsuperscript{1}\footnotemark[1],\\
{\textsuperscript{1} The Hong Kong University of Science and Technology} \\
{\textsuperscript{2} The Hong Kong University of Science and Technology (Guangzhou)} \\
{\textsuperscript{3} Peking University}\\
{\tt zkouaa@connect.ust.hk \{siruihan, yikeguo\}@ust.hk}\\
}

\begin{document}
\maketitle
\begin{abstract}
We present a novel three-stage framework leveraging Large Language Models (LLMs) within a risk-aware multi-agent system for automate strategy finding in quantitative finance. Our approach addresses the brittleness of traditional deep learning models in financial applications by: employing prompt-engineered LLMs to generate executable alpha factor candidates across diverse financial data, implementing multimodal agent-based evaluation that filters factors based on market status, predictive quality while maintaining category balance, and deploying dynamic weight optimization that adapts to market conditions. Experimental results demonstrate the robust performance of the strategy in Chinese \& US market regimes compared to established benchmarks. Our work extends LLMs capabilities to quantitative trading, providing a scalable architecture for financial signal extraction and portfolio construction. The overall framework significantly outperforms all benchmarks with 53.17\% cumulative return on SSE50~\footnote{The SSE 50 Index tracks the performance of the 50 most influential large-cap blue-chip stocks on the Shanghai Stock Exchange.} (Jan 2023 to Jan 2024), demonstrating superior risk-adjusted performance and downside protection on the market. Our protocol code are available at \url{https://github.com/kouzhizhuo/Automate-Strategy-Finding-with-LLM-in-Quant-investment}.
\end{abstract}
\section{Introduction}

Recent advances in LLMs and multi-agent systems are converging to transform quantitative finance. This synergistic relationship leverages LLMs' text comprehension and generation capabilities alongside multi-agent frameworks that simulate market dynamics, creating sophisticated approaches to portfolio management~\cite{Lee_2020}. 

LLMs have evolved from supporting analytical tools to active participants in financial decision-making~\cite{luo2025largelanguagemodelagent}. For example, BloombergGPT demonstrates superior performance in parsing market sentiment and answering domain-specific questions~\cite{wu2023bloomberggptlargelanguagemodel}. Research shows that LLMs effectively generate trading actions by contextualizing price trends with news and earnings reports~\cite{ding2023integratingstockfeaturesglobal}. Concurrently, multi-agent systems offer powerful approaches to portfolio optimization through decentralized interaction. The Multi-Agent Portfolio System demonstrates 
the portfolio managed by agents have achieved
well-diversified returns with improved risk-adjusted performance~\cite{Lee_2025}. These frameworks capture complex market dynamics such as information sharing and strategic arbitrage that single-agent models cannot address~\cite{spooner2020robustmarketmakingadversarial}. The integration of these technologies creates sophisticated financial environments where LLM-enhanced agents demonstrate adaptive behavior. StockAgent exemplifies this approach with LLM-powered agents mimic diverse investor personas responding to market events~\cite{zhang2024aimeetsfinancestockagent}. Hierarchical structures such as FinCon organize agents in manager-analyst relationships, facilitating collaboration through natural language communication~\cite{yu2024finconsynthesizedllmmultiagent}. This convergence heralds a future of intelligent, distributed financial decision making that combines data-driven learning with human-like reasoning capabilities for more robust investment strategies~\cite{yu2023finmemperformanceenhancedllmtrading}.

Alpha mining is the process of discovering trading signals that generate excess returns in financial markets, we identify three critical challenges in alpha mining: Rigidity of traditional methods that lack adaptability to dynamic markets~\cite{tang2025alphaagentllmdrivenalphamining};  Data diversity and integration challenges despite machine learning advances~\cite{cui2021alphaevolve,yu2023generating}; and Adaptation to market variability despite progress in prediction~\cite{xu2018stock} and strategy formulation~\cite{chen2019investment}.

Our LLM-driven framework addresses these limitations through three components: Flexible Alpha Mining employs LLMs to extract, categorize, and filter alpha factors from financial literature, organizing them as momentum, fundamental, or liquidity factors with established independence~\cite{xu2018stock}. Multi-agent Multimodal Market Evaluation conducts rigorous backtesting across diverse market conditions, with specialized agents evaluating factor effectiveness from multiple perspectives. Dynamic Strategy Optimization implements a weight gating layer that assigns optimal alpha factor weights based on current market conditions, ensuring adaptive strategy development.

Our methodology synthesizes cutting-edge machine learning techniques with established financial domain knowledge to create a robust interdisciplinary framework for alpha identification and optimization across diverse asset classes. This research is grounded in empirical quantitative investment practices, bridging theoretical advancements with practical applications in portfolio management.



Our main contributions are three-fold:
\ding{182} A novel framework for identifying formulaic alpha factors using LLMs, leveraging their exploratory capabilities to establish an Alpha factory from multimodal information with incremental update functionality;
\ding{183} Introduction of a multi-agent approach to portfolio management for evaluating relationships between market conditions and alpha factors, enabling specialized evaluation under different scenarios;
\ding{184} Integration of advanced techniques from machine learning and finance, representing a significant advancement in developing robust, adaptive investment strategies without human intervention.

The proposed framework demonstrates versatility across various asset classes, enhancing its utility and practical effectiveness. To support future research and ensure reproducibility, we make source code publicly available at~\url{https://github.com/kouzhizhuo/Automate-Strategy-Finding-with-LLM-in-Quant-investment}.

\section{Problem Formulation}
\label{sec:problem_formulation}

This section establishes the theoretical foundation for our research on alpha factor strategies in quantitative finance. We formulate a framework addressing three interconnected challenges: mathematically formalizing these alphas, developing dynamic methodologies to generate seed alphas, and defining and optimizing alpha factor strategies. Our approach integrates LLMs and multi-agent systems to overcome limitations in traditional quantitative trading methods.

We employ consistent notation throughout: $n$ stocks observed over trading periods $t \in {1, 2, \ldots, T}$, each characterized by $m$ financial features. Alpha factors are denoted as $\alpha_{ij}^{(t)}$ for stock $i$ in category $j$ at time $t$, with corresponding weights $w_i$. Market conditions at time $t$ are represented by $\mathcal{M}^{(t)}$, and alpha factor predictive power is measured using the Information Coefficient (IC)~\ref{IC}.

\begin{equation}
I C=\sigma(u, v)
\label{IC}
\end{equation}

A higher IC indicates stronger predictive relationships between alpha values and future returns. where $\sigma(u, v)$ is correlation coefficient between predicted alphas $u$ and actual future returns $v$.

\subsection{Alpha Factor Representation}
\label{subsec:alpha_representation}
The first challenge involves effectively representing alpha factors using mathematical expressions that capture meaningful financial signals~\cite{lo2007alphas}. Given raw financial features $\mathbf{X}i^{(t)} = \{X{i1}^{(t)}, X_{i2}^{(t)}, \ldots, X_{im}^{(t)}\}$ for each stock $i$ at time $t$ (such as price, volume, and volatility), we define alpha factors through two classes of operators:   

\textbf{Cross-Section Operators} $f_{\text{cs}}$: These operators process data from a single time period, capturing instantaneous relationships between financial variables:
\begin{equation}
\alpha_{cs}^{(t)} = f_{\text{cs}}(\mathbf{X}i^{(t)})
\end{equation}

\textbf{Time-Series Operators} $f_{\text{ts}}$: These operators analyze data spanning multiple periods, identifying trends and temporal patterns:
\begin{equation}
\alpha_{ts}^{(t)} = f_{\text{ts}}(\mathbf{X}i^{(t)}, \mathbf{X}_i^{(t-1)}, \ldots, \mathbf{X}_i^{(t-n)})
\end{equation}

The seed alpha for stock $i$ in category $j$ at time $t$ is formulated as:
\begin{equation}
\alpha_{ij}^{(t)} = w_{cs} \cdot f_{\text{cs}}(\mathbf{X}i^{(t)}) + w{ts} \cdot f_{\text{ts}}(\mathbf{X}i^{(t)}, \ldots, \mathbf{X}_i^{(t-n)})
\end{equation}
\noindent where $w_{cs}$ and $w_{ts}$ are weights assigned to the cross-section and time-series components respectively. This representation allows for the flexible combination of different financial signals, enabling the creation of complex and nuanced alpha factors.
\subsection{Alpha Mining and Selection}
\label{subsec:alpha_mining}
The second challenge addresses the limitations of traditional alpha mining methods, which often fail to adapt to changing market conditions. We formulate a dynamic selection approach that identifies the most relevant alphas for current market conditions using two complementary evaluation mechanisms: 

\textbf{Confidence Score Evaluation}: Assesses the statistical reliability of each alpha factor:
\begin{equation}
\theta_{ij} = \mathbb{E}\left[IC(\alpha_{ij}^{(t)} | \mathcal{M}^{(t)})\right]
\end{equation}
where $\theta_{ij}$ represents the confidence score for alpha $\alpha_{ij}$, and $\mathbb{E}[\cdot]$ denotes the expected value. Higher confidence scores indicate more consistent performance across various market environments.

\textbf{Risk Preference Evaluation}: Examines the risk characteristics of each alpha factor:
\begin{equation}
\rho_{ij} = f_{\text{risk}}(\alpha_{ij}^{(t)}, \mathcal{M}^{(t)})
\end{equation}
where $\rho_{ij}$ represents the risk score and $f_{\text{risk}}$ is a function that evaluates how well the alpha performs under different risk scenarios.

The optimal set of seed alphas is selected by considering both confidence and risk evaluations:
\begin{equation}
\alpha_{ij}^{} = \underset{\alpha{ij}}{\arg\max} \, \left[ w_c \cdot \theta_{ij} + w_r \cdot \rho_{ij} \right]
\end{equation}
where $w_c$ and $w_r$ are weights assigned to confidence and risk scores respectively, reflecting their relative importance in the selection process.
\subsection{Strategy Optimization}
\label{subsec:strategy_optimization}
The third challenge involves combining the selected alphas into an effective investment strategy. The final alpha strategy is defined as a weighted combination of the optimal alphas from each category:
\begin{equation}
\alpha^{(t)} = \sum_{j=1}^{k} w_j \cdot \alpha_{ij}^{}
\end{equation}
where $w_j$ represents the weight assigned to category $j$, and $k$ is the total number of alpha categories.
These weights are dynamically adjusted based on current market status to maximize strategy performance while managing overall portfolio risk. This optimization completes our framework, transforming raw financial data into a robust trading strategy that can adapt to changing market environments.

\begin{figure*}[!t]
    \centering
    \includegraphics[width=.9\linewidth]{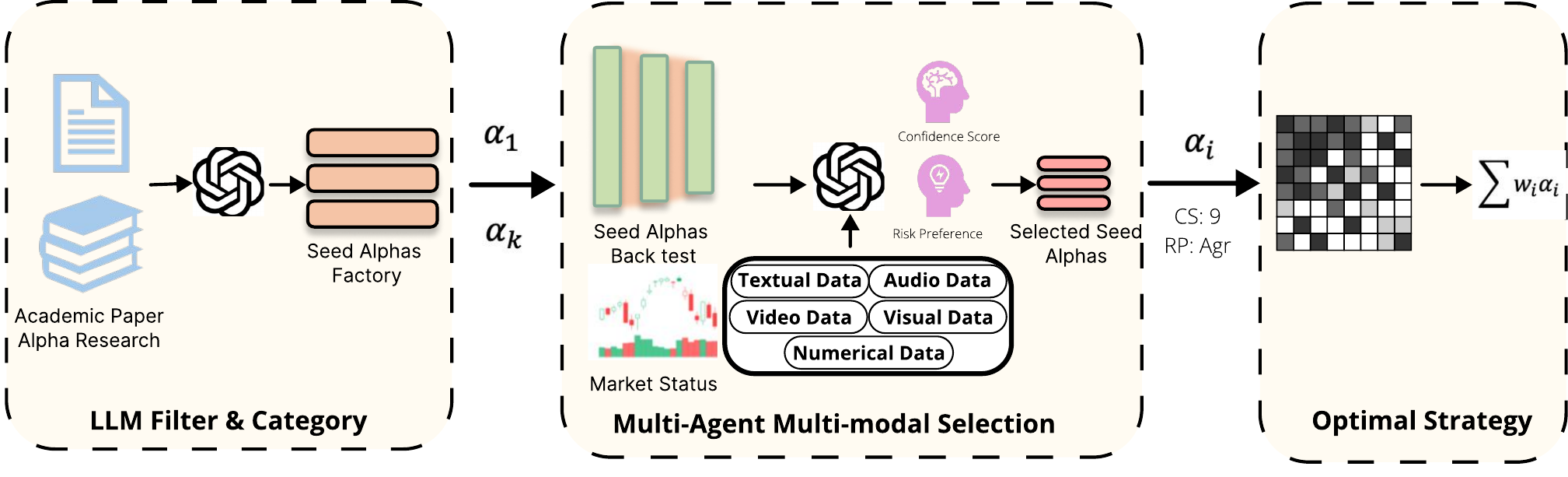}
    \caption{Overview of the strategy generate process in three components, a Seed Alpha Factory built using LLMs, a multi-agent decision-making system, and a weight optimization approach for overall strategy (CS stands for confidence score; RP stands for risk preference)}
    \label{methodology}
\end{figure*}

\section{Methodology}
\subsection{Framework Overview}
\label{subsec:framework_overview}

Our framework comprises three interconnected components (Figure~\ref{methodology}): the SAF, multi-agent decision-making, and weight optimization. The initial phase employs LLMs to analyze and categorize multimodal financial research documents, constructing a comprehensive SAF. The LLM's capability to process diverse datasets enables the creation of a robust set of seed alphas categorized into independent groups, aligning with established finance alpha mining principles~\cite{openai2023gpt4}. The second phase implements a multimodal multi-agent evaluation process that incorporates varied risk perspectives, enhancing strategy adaptability across different market conditions. This phase produces an optimized alpha set tailored to current market states and risk preferences. The final phase employs deep learning methods to optimize the weights of selected alphas, constructing a cohesive overall strategy.

The framework's dynamic architecture enables continuous refinement through incremental updates to the SAF as new research emerges and market conditions evolve. This adaptability ensures the strategy maintains relevance and robustness over time. The methodology's versatility permits application to any structured market globally, effectively replicating and enhancing professional investment research approaches.

\subsection{LLM-Based Seed Alpha Generation}
\label{subsec:llm_seed_alpha}

The first stage implements an LLM-based filtering and categorization process for alpha-related research. We utilize GPT-4o model to perform these tasks on a diverse corpus of financial research. Our initial dataset comprises 11 documents spanning both theoretical and applied aspects of alpha mining research (see Appendix~\ref{sec:appendix 1}). Through this process, the system generates 9 distinct categories containing 100 seed alphas. The approach enhances the model's ability to extract intricate details and relationships within financial research, resulting in a more robust and diverse SAF.

The output is a structured set of seed alphas categorized into distinct financial domains such as Momentum, Mean Reversion, and Fundamental analysis et al. Each category includes specific alpha designations and corresponding executable formulations derived from the LLM's analysis (see Appendix~\ref{sec:appendix 2}). This structured output forms the foundation for subsequent processing stages.

\subsection{Multimodal Multi-Agent Evaluation}
The second stage implements a comprehensive evaluation and selection of alpha factors through a multimodal and multi-agent system. Our system incorporates five types of multimodal data (detailed in Appendix~\ref{sec:appendix 3}) encompassing textual, numerical, visual, audio, and video inputs. The multi-agent architecture comprises two agents Confidence Score Agent (CSA) and Risk Preference Agent (RPA). Selected alpha factors undergo rigorous backtesting using historical market data, with key evaluation metrics including the IC~\cite{goodwin1998information}~\ref{IC} and Sharpe Ratio~\cite{sharpe1994sharpe}~\ref{sharpe ratio}:
\begin{equation}
\text{Sharpe Ratio} = \frac{\mathbb{E}[R(\alpha^{(t)}) - R_f]}{\sqrt{\text{Var}[R(\alpha^{(t)})]}}
\label{sharpe ratio}
\end{equation}
where $R(\alpha^{(t)})$ represents the return of the strategy and $R_f$ represents the risk-free rate.
We developed a Category-Based Alpha Selection algorithm (detailed in Appendix~\ref{sec:appendix 4}) to automate the selection. This algorithm systematically identifies and selects alphas from different categories based on their confidence scores, ensuring rigorous selection of factors that meet confidence thresholds across all categories while maintaining category diversity.
\subsection{Weight Optimization}
\label{subsec:weight_optimization}

The final stage employs a 3-layer MLP to optimize the weights of selected seed alphas by mapping historical alpha calculations to future yields~\cite{chen2020dynamic}. The architecture consists of an input layer processing daily alpha values, a hidden layer with ten ReLU-activated nodes, and a single-node output layer for yield prediction.

During training, the network employs backpropagation and gradient descent to minimize the loss function, quantifying the discrepancy between predicted and actual yields. We utilize a separate validation set to ensure model generalizability and prevent overfitting. The DNN processes input data through the hidden layer, transforming it with learned weights and biases, and generates the final output through the output layer's weights, biases, and activation function.

This methodology establishes a robust framework for predicting future yields based on historical alpha values, demonstrating the efficacy of deep learning techniques in optimizing alpha weights and enhancing investment strategy performance. Algorithm~\ref{overallago} provides a formal specification of our complete framework, illustrating the logical flow from multimodal data processing through multi-agent evaluation to final weight optimization.

\section{Experiment}
Our research aims to develop a comprehensive LLM-driven alpha mining framework that operates without human intervention. This framework is uniquely capable of processing multimodal information and adapting to varying market conditions. To validate the effectiveness of our framework, we have conducted a series of experiments.


\subsection{Datasets}
Our study focuses on financial data from the Chinese market and US market, specifically targeting the SSE 50 Index. Table~\ref{Dataset1} shows the experiment dataset, which encompasses six primary features as original inputs for our Alpha factors: open, high, low, close, volume (OHLCV), and volume-weighted average price (VWAP). To ensure rigorous evaluation and robust model performance. Our experiments integrate financial reports and factor performances of the 50 constituent companies of the SSE 50 Index, providing a comprehensive view of the market~\cite{li2020asset}. The evaluation considers various datasets, including financial reports from the specified periods and performance metrics of different Alpha factors. 
\begin{table}[!t]
    \centering
    \caption{Summary of the experiment dataset}
    \resizebox{\linewidth}{!}{
    \begin{tabular}{c|c }\hline
\bf Aspect&\bf Details\\ \hline
Primary Features&Open, High, Low, Close, Volume, VWAP\\
Alpha Factors&Custom factors based on price, volume, financial ratios,\\
& moving averages, sentiment analysis\\
Financial Reports&Quarterly and Annual reports from Index constituent companies\\
Time Periods& Jan 2019-Jun 2024\\
Market Coverage&SSE50, CSI300, SP500 Index\\
Evaluation Criteria&Causal relationships, Alpha factor performance, model robustness\\ \hline
    \end{tabular}
    \label{Dataset1}}
\end{table}

\subsection{Multimodal Knowledge Extraction and Adaptive Alpha Discovery}

\begin{figure}[!t]
    \centering
    \includegraphics[width=1\linewidth]{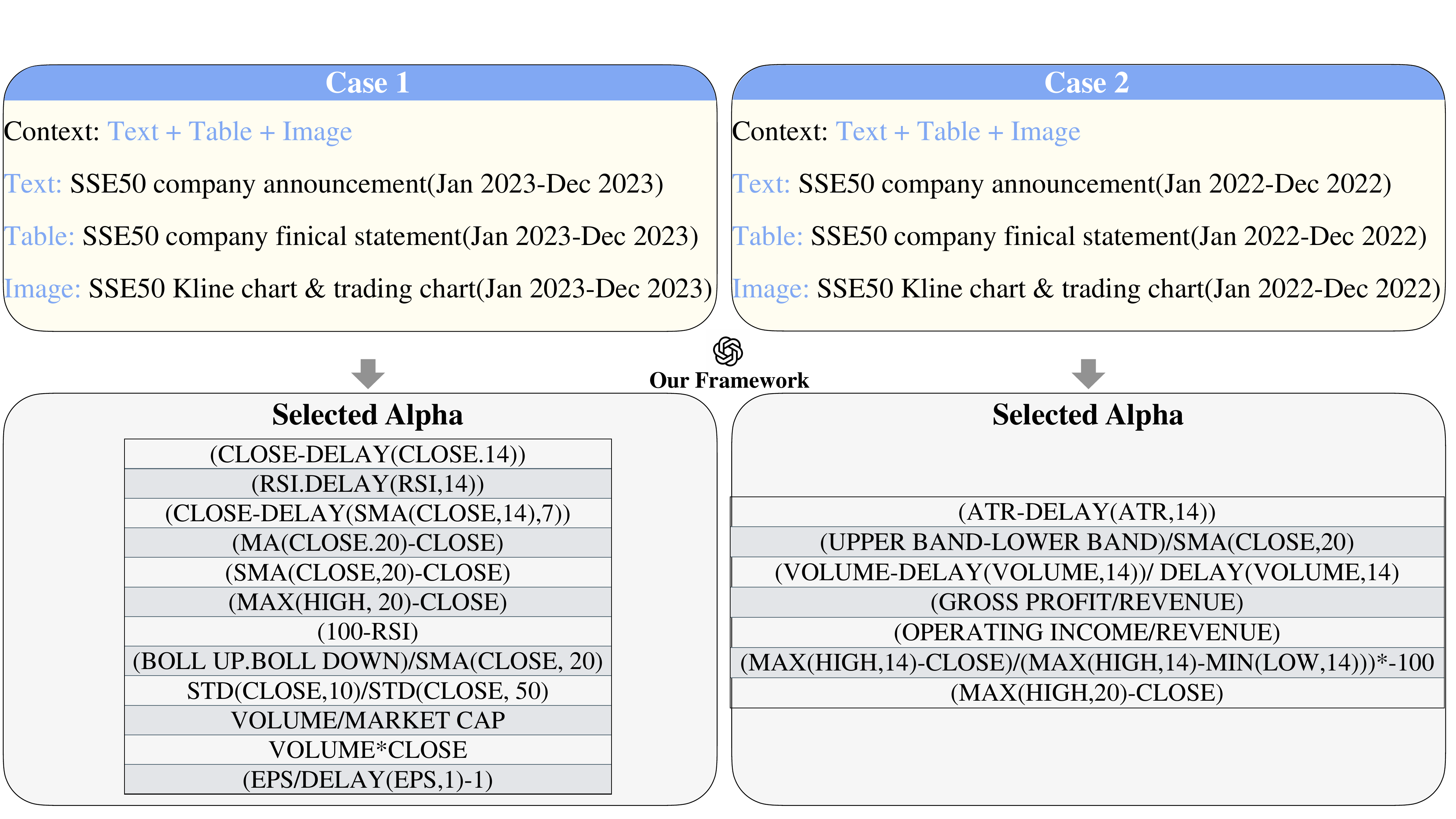}
    \caption{Sample Experiment on Different Market Status Input and Alpha Selection, Selected Alpha Depends on Different Context}
    \label{prompt}
\end{figure}

We implement a prompt architecture (Figure~\ref{prompt}) that incorporates multimodal market information into LLMs for comprehensive knowledge extraction and seed alpha selection. This architecture integrates textual financial sentiment data, numerical company financial statements, and visual trading charts to provide holistic market analysis~\cite{luo2025sparsecausaldiscoverygenerative}. Our contextual analysis mechanism dynamically adjusts parameters based on prevailing market trends and sector performance. Experimental validation demonstrates the framework's adaptability across varying market conditions. In Case 1, analyzing SSE50 data from Jan 2023 to Dec 2023, the model selected momentum and volume-based indicators such as price momentum, RSI, and MACD. Conversely, in Case 2, when processing the time window from Jan 2022 to Dec 2022, the model prioritized volatility and economic factors, including ATR, Bollinger Bands, and gross profit indicators.

These results confirm the framework's capacity to dynamically adapt to changing market conditions through effective multimodal data integration. This adaptability enables the identification of market-appropriate alphas, enhancing strategy robustness across diverse market environments.

\subsection{Comparative Performance Against Traditional Alpha Factories}
When evaluating the performance of selected seed alpha signals, the primary metric is the IC. We evaluated five most common alpha categories: Momentum, Mean Reversion, Volatility, Fundamental and Growth. The results in Table~\ref{tab:ic_comparison} demonstrate that our LLM-driven framework consistently achieves higher average IC values across all categories, particularly in Volatility and Fundamental, indicating superior trading effectiveness compared to original.


\begin{table}[t]
  \centering
  \tabcolsep 0.1cm
  \caption{IC comparison of mean \& selected}
  \resizebox{\linewidth}{!}{
  \begin{tabular}{m{3cm}<\centering ccccc}
    \toprule
    & Momentum & \makecell[c]{Mean\\ Reversion} & Volatility & Fundamental & Growth \\
    \midrule
    Mean IC of SAF & 0.0092 & 0.0135 & 0.0177 & 0.0118 & 0.0146 \\
    Mean IC of Selected SAF & \textbf{0.0208} & \textbf{0.0187} & \textbf{0.0258} & \textbf{0.0192} & \textbf{0.0217} \\
    \bottomrule
  \end{tabular}}
  \label{tab:ic_comparison}
\end{table}

\subsection{Performance Evaluation of the Integrated Alpha Framework in Benchmark Comparison}
\label{4.4}
\begin{figure*}[t]
    \centering
    \includegraphics[width=0.8\linewidth,height=5cm]{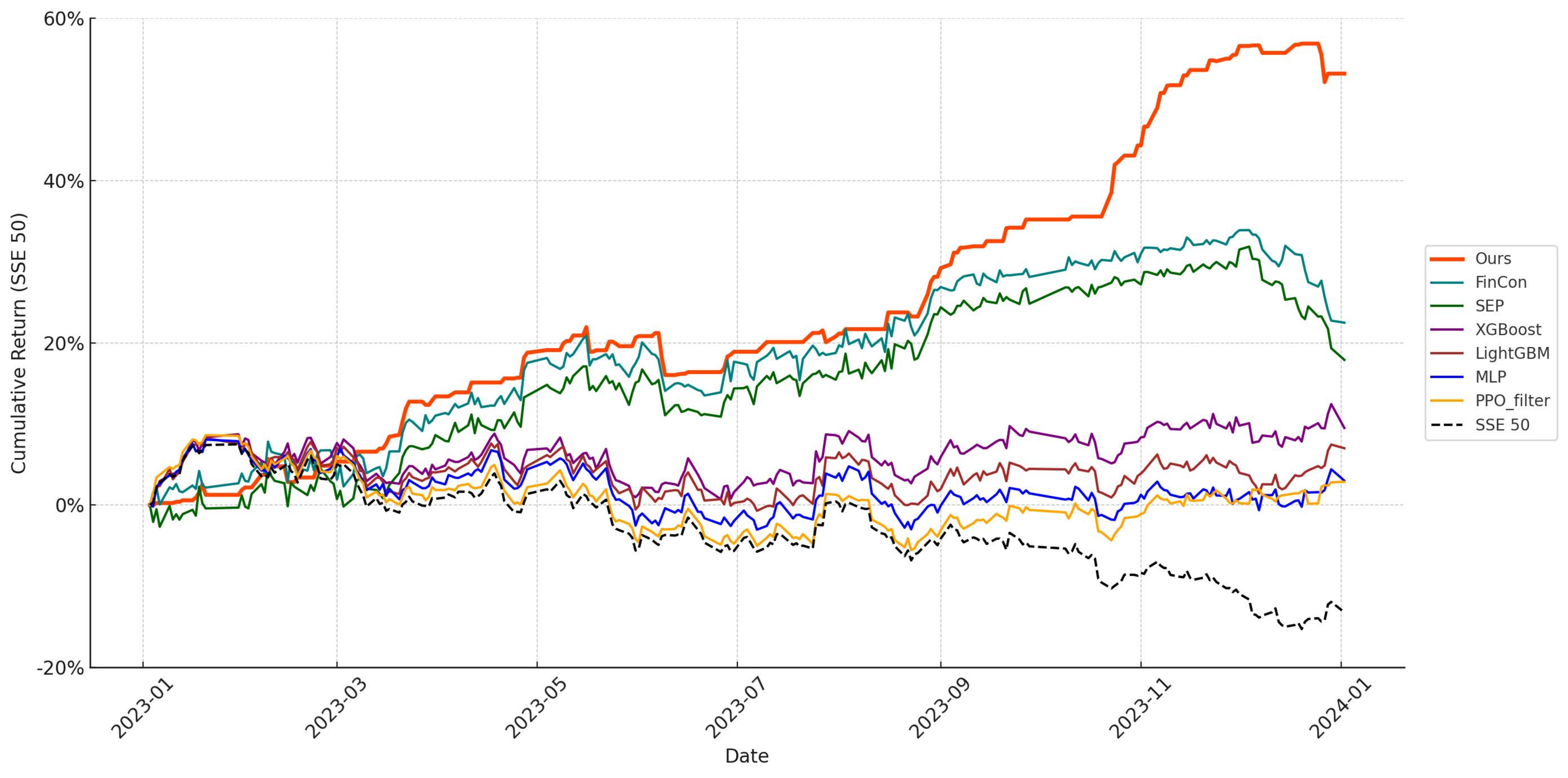}
    \caption{Cumulative return backtest result on SSE50. The line track the net worth of different methods}
    \label{main result}
\end{figure*}

\begin{table}[htbp]
    \centering 
    \caption{SSE50 2023 test combination of 12 Alphas} 
\resizebox{\linewidth}{!}{
\begin{tabular}{c|c|c|c}\toprule
$\#$ & Alpha & Weight & IC(SSE50) \\ \midrule
 1 & (CLOSE - DELAY(CLOSE, 14)) & -0.1459 & \cellcolor{gray!40} 0.0209 \\
 2 & (RSI - DELAY(RSI, 14)) & -1.0265 &\cellcolor{gray!35} -0.0225 \\
 3 & (CLOSE - DELAY(SMA(CLOSE, 14), 7)) & -0.1978 &\cellcolor{gray!45} 0.0193 \\
 4 & (MA(CLOSE, 20) - CLOSE) & 0.0556 &\cellcolor{gray!75} -0.0186 \\
 5 & (SMA(CLOSE, 20) - CLOSE) & -0.945 &\cellcolor{gray!75} -0.0186 \\
 6 & (MAX(HIGH, 20) - CLOSE) & -0.4053 &\cellcolor{gray!70} -0.0185 \\
 7 & (100-RSI) & -0.3199 &\cellcolor{gray!45} 0.0194 \\
 8 & (BOLL\_UP - BOLL\_DOWN) / SMA(CLOSE, 20) & 3.6186 &\cellcolor{gray!25} 0.0278 \\
 9 & STD(CLOSE, 10) / STD(CLOSE, 50) & -0.183 &\cellcolor{gray!30} 0.0236 \\
 10 & VOLUME / MARKET\_CAP & -3.2145 &\cellcolor{gray!60} -0.0194 \\
 11 & VOLUME * CLOSE & -0.0058 &\cellcolor{gray!50} 0.0187 \\
 12 & (EPS / DELAY(EPS, 1) - 1) & -1.8351 & \cellcolor{gray!65}-0.0215 \\  \midrule
 &\multicolumn{2}{|c}{Weighted Combination}&\cellcolor{gray!80} -0.0587 \\ \bottomrule
\end{tabular}}
\label{selected alpha SSE50}
\end{table}

Table~\ref{selected alpha SSE50} presents an example combination of 12 alphas generated by our framework, evaluated on the SSE50 constituent stock set. The table details the seed alphas selected by the LLM from each category, along with their respective weights and IC~\ref{IC} values. The weight combination IC~\ref{IC} value is quite high as -0.0587. Although some of the seed alphas exhibit relatively low IC~\ref{IC} values individually, their removal results in a significant drop in the retrained combination weight, indicating their critical role in the overall performance. For example, if we remove alpha \#6 the weight combination will drop to -0.055; once we remove the alpha \#11, the weight combination will drop to merely 0.0491. This suggests that the seed alpha set selected by the LLM synergizes effectively, providing robust predictive power~\cite{zhang2020information}.  To address the question of whether our overall strategy, incorporating the LLM-driven framework and confidence scoring, can consistently beat the market, we conducted a backtest using a straightforward investment approach during the period from Jan 2023 to Jan 2024 (training: Jan 2021-Jun 2022; validation: Jul 2022-Dec 2022), on the SSE50 dataset we conduct the framework to construct the portfolio (~\ref{portfolio construction}). The net worth progression of the respective strategies over the testing period is shown in Figure~\ref{main result}, and the comparison matrix shows in Table~\ref{strategy_comparison matrix}. Despite not explicitly optimizing for absolute returns, our framework demonstrates impressive performance in the backtest, achieving the highest profit compared to other methods. Our cumulative return for 2023 backtest comes to $53.17 \%$ positive and meanwhile the index performance is -11.73\%. We also do the compression with 4 advanced statistical methods and 2 LLM based methods. The framework strategy substantially outperforms all competitors with the highest return (53.17\%), best risk-adjusted metrics (Sharpe: 0.287, Sortino~\footnote{ Sortino ratio computing excess return per unit of downside deviation, evaluating only negative return volatility.}: 0.208, Calmar~\footnote{Calmar ratio divides annualized return by maximum drawdown, measuring reward relative to risk.}: 1.052), and lowest volatility (0.762\%). LLM based mathods FinCon and SEP show moderate success (22.47\%, 17.89\%), while conventional machine learning methodologies yield marginal returns (XGBoost 9.53\%). Notably, all evaluated strategies surpass the SSE50 benchmark, which exhibits negative performance (-13.22\%). This approach demonstrates the efficacy of our LLM-driven framework consistently outperforming market benchmarks through dynamic adaptation to changing conditions.

\begin{table}[t]
  \centering
  \tabcolsep 0.1cm
  \caption{Performance comparison of trading strategies}
  \resizebox{\linewidth}{!}{
  \label{strategy_comparison matrix}
  \begin{tabular}{p{3cm}<\raggedright ccccc}
    \toprule
    \textbf{Strategy} & \makecell[c]{\textbf{Final}\\ \textbf{Return (\%)}} & \makecell[c]{\textbf{Sharpe}\\ \textbf{Ratio}} & \makecell[c]{\textbf{Volatility}\\ \bf (\%)} & \makecell[c]{\textbf{Sortino}\\ \bf Ratio} & \makecell[c]{\textbf{Calmar}\\ \bf Ratio} \\
    \midrule
    Ours & \textbf{53.173} & \textbf{0.287} & \textbf{0.762} & \textbf{0.208} & \textbf{1.052} \\
    XGBoost~\cite{chen2016xgboost} & 9.532 & 0.038 & 1.019 & 0.067 & 0.103 \\
    LightGBM~\cite{ke2017lightgbm} & 7.125 & 0.030 & 0.993 & 0.053 & 0.066 \\
    MLP & 3.110 & 0.013 & 0.960 & 0.023 & 0.043 \\
    PPO\_filter~\cite{schulman2017proximalpolicyoptimizationalgorithms} & 2.865 & 0.013 & 0.886 & 0.024 & 0.017 \\
    FinCon~\cite{yu2024finconsynthesizedllmmultiagent} & 22.474 & 0.077 & 1.196 & 0.126 & 0.232 \\
    SEP~\cite{koa2024learning} & 17.891 & 0.060 & 1.217 & 0.103 & 0.157 \\
    SSE 50 & -13.22 & -0.063 & 0.859 & -0.111 & -0.043 \\
    \bottomrule
  \end{tabular}}
\end{table}

\subsection{Framework Robustness Across Time Periods and Markets}
We conducted rigorous cross-temporal and cross-market validation using two comprehensive datasets: CSI300~\footnote{CSI300 Index tracking the 300 largest companies listed on China's Shanghai and Shenzhen exchanges.} constituents (accessed via Tushare API) and SP500~\footnote{SP500 Index measuring the performance of 500 large U.S. companies traded on American stock exchanges.} constituents (accessed via CRSP platform) spanning January 2019 to December 2023. Our dataset incorporated daily OHLCV price metrics, quarterly financial statements, and relevant macroeconomic indicators to enable robust alpha factor construction. To ensure methodological integrity, we systematically partitioned the dataset into training, validation, and testing subsets with precise chronological segmentation as detailed in Table~\ref{test time}. This experimental design facilitates comprehensive evaluation of our framework's generalizability across different market environments and temporal contexts.

\begin{table}[htbp]
    \centering
    \caption{Training, validation, and test Periods for CSI300 and SP500}
    \resizebox{\linewidth}{!}{
    \begin{tabular}{l|c|c|c}
        \hline
        \textbf{Assets} & \textbf{Training Period} & \textbf{Validation Period} & \textbf{Test Period} \\
        \hline
        \multirow{3}{*}{CSI300} & Jan 2019-Jun 2020 & Jun-Dec 2020 & Jan-Jun 2021 \\
        & Jan 2020-Jun 2021 & Jun-Dec 2021 & Jan-Jun 2022 \\
        & Jan 2021-Jun 2022 & Jun-Dec 2022 & Jan-Jun 2023 \\
        \hline
        \multirow{3}{*}{SP500} & Jan 2019-Jun 2020 & Jun-Dec 2020 & Jan-Jun 2021 \\
        & Jan 2020-Jun 2021 & Jun-Dec 2021 & Jan-Jun 2022 \\
        & Jan 2021-Jun 2022 & Jun-Dec 2022 & Jan-Jun 2023 \\
        \hline
    \end{tabular}}
    \label{test time}
\end{table}

We employed consistent temporal partitioning across Chinese A-share and US markets to enable direct comparative analysis. Our findings demonstrate the framework's robust performance across diverse market environments. From Table~\ref{backtest 2} shows in the Chinese A-Share market, the strategy generated substantial alpha with CSI300 constituents, particularly in H1 2023 (192.27\% annual return vs. benchmark 9.13\%). Similarly, in the US market, the framework achieved strong returns with SP500 constituents (93.61\% in H1 2021, 118.24\% in H1 2023). Notably, during the H1 2022 market downturn, the strategy maintained positive returns in both markets (12.78\% in CSI300, 2.77\% in SP500) while their respective benchmarks declined significantly (-30.37\% and -44.22\%).

\begin{table}[t]
    \centering 
    \tabcolsep 0.1cm
    \caption{Backtest performance results across different time windows}
    \label{tab:backtest_results}
    \resizebox{\linewidth}{!}{
    \begin{tabular}{l|cc|cc|cc}\toprule
    \textbf{Time Window} & \multicolumn{2}{c|}{\textbf{Annual Return (\%)}} & \multicolumn{2}{c|}{\textbf{Cum. Return (\%)}} & \multicolumn{2}{c}{\textbf{Max DD (\%)}} \\
     & \textbf{Strategy} & \textbf{Baseline} & \textbf{Strategy} & \textbf{Baseline} & \textbf{Strategy} & \textbf{Baseline} \\ \midrule
    \multicolumn{7}{c}{\textit{Our Strategy vs CSI300}} \\ \midrule
    Jan-Jun 2021 & \textbf{29.70} & 7.31 & \textbf{11.91} & 3.10 & -19.40 & \textbf{-10.86} \\
    Jan-Jun 2022 & \textbf{12.78} & -30.37 & \textbf{5.29} & -14.37 & \textbf{-24.01} & -28.59 \\
    Jan-Jun 2023 & \textbf{192.27} & 9.13 & \textbf{59.03} & 3.85 & -17.03 & \textbf{-8.44} \\ \midrule
    \multicolumn{7}{c}{\textit{Our Strategy vs SP500}} \\ \midrule
    Jan-Jun 2021 & \textbf{93.61} & 29.67 & \textbf{35.19} & 12.59 & -7.89 & \textbf{-4.23} \\
    Jan-Jun 2022 & \textbf{2.77} & -44.22 & \textbf{1.25} & -23.39 & \textbf{-20.55} & -23.51 \\
    Jan-Jun 2023 & \textbf{118.24} & 35.22 & \textbf{42.78} & 14.76 & -11.52 & \textbf{-7.75} \\ \bottomrule
    \multicolumn{7}{p{12cm}}{\small \textit{Note}:Max DD = Maximum Drawdown.} \\
    \end{tabular}}
    \label{backtest 2}
\end{table}

The empirical evidence substantiates both the temporal persistence and cross-market robustness of our findings, with notably beneficial countercyclical characteristics emerging during periods of market distress. The framework demonstrates consistent alpha generation across diverse market architectures, regulatory frameworks, and investor behavioral patterns, establishing its broad applicability. Specifically, the model maintains stable outperformance trajectories through bullish, bearish, and range-bound market conditions, thereby validating its structural integrity and adaptability to varying macroeconomic environments.

\subsection{Ablation Study}
\label{sec:ablation_study}
Our ablation study systematically evaluates the contribution of CSA and RPA within a multi-agent framework. We evaluate three configurations of each, using the SSE50 index data (2010-2022) and measure performance by IC~\ref{IC} and Sharpe Ratio~\ref{sharpe ratio}.

Table \ref{tab:ablation_results} demonstrates that the complete model achieved superior performance, with an out-of-sample IC of 0.047 and Sharpe Ratio of 1.73. Removing the CSA caused substantial degradation, reducing out-of-sample IC by 31.9\% and Sharpe Ratio by 22.5\%. RPA removal also decreased performance metrics, with particularly significant impact on the Sharpe Ratio. These results indicate that while both components contribute meaningfully, the CSA plays a more critical role in maintaining predictive stability. To assess performance consistency across market conditions, we analyzed out-of-sample IC values during different market regimes (Table \ref{tab:market_regimes}). The complete model maintained consistent performance across bull, bear, and sideways markets. In contrast, the model without CSA performed particularly poorly during bear markets(IC: 0.021 vs. 0.042). The model without RPA showed moderate degradation during non-bull markets, less severe than observed without CSA.

This analysis confirms that both components enhance performance stability, with the CSA providing especially critical functionality during adverse market conditions.

\begin{table}[!t]
    \centering
    \caption{Ablation study results: impact of agent components on performance metrics}
    \resizebox{\linewidth}{!}{
    \begin{tabular}{lccc}\hline
        \textbf{Model Configuration} & \makecell[c]{\textbf{IC}\\ \bf (In-Sample)} & \makecell[c]{\textbf{IC}\\ \bf (Out-of-Sample)} & \makecell[c]{\textbf{Sharpe}\\ \bf Ratio} \\ \hline
        Full Model & \textbf{0.059} & \textbf{0.047} & \textbf{1.94} \\
        Without Confidence Score & 0.054 & 0.032 & 1.51 \\
        Without Risk Preference & 0.056 & 0.039 & 1.34 \\ \hline
    \end{tabular}}
    \label{tab:ablation_results}
\end{table}

\begin{table}[!t]
    \centering
    \caption{Performance across market regimes (out-of-sample IC)}
    \resizebox{\linewidth}{!}{
    \begin{tabular}{lccc}\hline
        \textbf{Model Configuration} & \makecell[c]{\textbf{Bull}\\ \bf Market} & \makecell[c]{\textbf{Bear}\\ \bf Market} & \makecell[c]{\textbf{Sideways}\\ \bf Market} \\ \hline
        Full Model & \textbf{0.051} & \textbf{0.042} & \textbf{0.045} \\
        Without Confidence Score & 0.046 & 0.021 & 0.029 \\
        Without Risk Preference & 0.049 & 0.028 & 0.037 \\ \hline
    \end{tabular}}
    \label{tab:market_regimes}
\end{table}

\subsection{Sensitive Study}

We analyzed the framework's sensitivity to agent weights and neural network hyperparameters. The CSA \& RPA weight ratio of 0.6/0.4 provided superior performance across market regimes, particularly during bear markets (Sharpe: 10.37), indicating enhanced robustness to regime shifts (Table \ref{tab:agent_weights}). Optimal neural network configuration (10 hidden nodes, learning rate: 0.001, batch size: 32, regularization: 0.001) achieved a Sharpe ratio of 13.33, with performance particularly sensitive to regularization strength (Table \ref{tab:nn_hyperparams}).

\begin{table}[t]
\centering
\caption{Sensitivity to agent weight configurations}
\label{tab:agent_weights}
\resizebox{\linewidth}{!}{
\begin{tabular}{ccccc}
\hline
Confidence & Risk & \multicolumn{3}{c}{Sharpe Ratio} \\
\ Weight & \ Weight & \ Bull Market & \ Bear Market & \ Overall \\
\hline
1.0 & 0.0 & 4.32 & 6.60 & 8.10 \\
0.8 & 0.2 & -3.41 & -7.23 & -2.68 \\
0.6 & 0.4 & \textbf{8.70} & \textbf{10.37} & \textbf{11.39} \\
0.5 & 0.5 & -0.49 & -1.62 & -5.10 \\
0.4 & 0.6 & -4.27 & -4.41 & -8.04 \\
0.2 & 0.8 & 5.68 & 8.51 & 9.00 \\
0.0 & 1.0 & -0.61 & 1.78 & -4.35 \\
\hline
\end{tabular}}
\end{table}

\begin{table}[t]
\centering
\caption{Sensitivity to neural network hyperparameters}
\label{tab:nn_hyperparams}
\resizebox{\linewidth}{!}{
\begin{tabular}{ccccc}
\hline
\ Hidden & \ Learning & \ Batch & \ Regularization & \ Sharpe \\
\ Nodes & \ Rate & \ Size & \ Parameter & \ Ratio \\
\hline
5 & 0.001 & 32 & 0.001 & 9.69 \\
10 & 0.001 & 32 & 0.001 & \textbf{13.33} \\
20 & 0.001 & 32 & 0.001 & 6.93 \\
10 & 0.0005 & 32 & 0.001 & 7.03 \\
10 & 0.002 & 32 & 0.001 & 5.13 \\
10 & 0.001 & 16 & 0.001 & 6.26 \\
10 & 0.001 & 64 & 0.001 & 4.25 \\
10 & 0.001 & 32 & 0.0005 & -1.88 \\
10 & 0.001 & 32 & 0.002 & -6.91 \\
\hline
\end{tabular}}
\end{table}

The analysis revealed regime-dependent behavior across market conditions, and critical sensitivity to regularization strength. The optimal configuration achieved a robust overall Sharpe ratio of 11.39.

\section{Related Work}

\textbf{Formulaic alphas} in quantitative investment represent systematic, rule-based strategies that generate excess returns by exploiting specific market patterns and inefficiencies \cite{kakushadze2016101}. These strategies employ various methodologies, including genetic programming that involves structural and numerical mutations to generate novel alphas \cite{cong2021alphaportfolio}, enhanced time-series operators with mutual information as fitness measures \cite{lin2019alpha}, and algorithmic graphs for more complex predictions \cite{cui2021alphaevolve}. Machine learning approaches utilize neural network architectures such as LSTM \cite{hochreiter1997lstm} and Transformer models \cite{vaswani2017attention}, while decision tree models like XGBoost \cite{chen2016xgboost} and LightGBM \cite{ke2017lightgbm} offer interpretability advantages. Recent research focuses on integrating non-standard data sources, as demonstrated by REST \cite{xu2021rest} and HIST \cite{xu2021hist}.

The development of \textbf{general-domain LLMs} has catalyzed interest in Finance LLMs (Fin-LLMs), although this specialized domain remains nascent \cite{novy-marx2015momentum, yang2023investlm, zhao2023survey}. Open-source LLMs such as LLaMA \cite{touvron2023llama}, BLOOM \cite{bloom2023}, and Flan-T5 \cite{chung2022scaling} provide flexibility but may underperform proprietary alternatives. Fine-tuned financial LLMs demonstrate enhanced domain-specific comprehension, yet their generative performance indicates the need for improved domain-specific datasets \cite{lewis2020retrieval, koa2024learning, wen2025thinkpatterns21ksystematicstudyimpact}.

\textbf{Multimodal LLMs} have shown significant potential in investment contexts by processing diverse data types \cite{li2023multimodal, luo2025finmmebenchmarkdatasetfinancial}, developing strategies that mitigate market volatility \cite{ouyang2024modal}, and analyzing textual data to gauge investor sentiment \cite{zhao2024revolutionizing}. Multi-agent LLM systems enhance market analysis capabilities by leveraging vast datasets to interpret financial reports and market sentiment \cite{zhang2024multimodal}, simulating various market scenarios \cite{talebirad2023multiagent}, and facilitating parallel testing of diverse strategies \cite{wang2024rethinking}. Implementation raises important considerations regarding transparency, accountability, and bias mitigation \cite{yu2024finmem, mundhenk2021symbolic}.

\section{Conclusion}

In this paper we proposed a novel quantitative investment framework integrating LLMs and multi-Agent architectures to address instability in traditional approaches. Our system generates diversified alpha factors from multimodal financial data, constructs risk-calibrated trading agents, and employs a deep learning mechanism for dynamic agent weighting based on market conditions. 

Experimental results confirm the framework's effectiveness across CN and US markets, our framework demonstrates significant outperformance versus SOTA alpha generation methods and benchmark indices across key financial metrics. This work successfully extends LLM capabilities to quantitative trading, creating a scalable, adaptive architecture for financial signal extraction that functions effectively without human intervention.

\section{Limitations}

Our framework presents several significant limitations. First, system efficacy is contingent upon input document quality, potentially perpetuating inherent biases~\cite{deb2017review, ashok2018online}. Second, LLM-generated alphas occasionally lack the financial intuition characteristic of human analysts, resulting in theoretically sound but practically infeasible factors~\cite{tuarob2017feeling}. Third, our multi-agent evaluation methodology presupposes persistent historical relationships between market conditions and alpha performance, an assumption that may prove tenuous during market regime shifts.  Finally, our validation efforts have primarily targeted equity markets, with cross-asset applicability requiring additional empirical investigation. Future research directions should address these constraints through exploring Mixture of Experts (MoE) architectures to improve learning efficiency~\cite{masoudnia2014mixture}, adaptive agent architectures, transfer learning methodologies~\cite{wang2023methods}, related regulation requirements~\cite{arner_financial_2020}, and computationally efficient implementations.

\section{Acknowledgments}

This paper is partially supported by grants from the HKUST Start-up Fund (R9911), Theme-based Research Scheme grant (T45-205/21-N), the InnoHK initiative of the Innovation and Technology Commission of the Hong Kong Special Administrative Region Government, and the research funding under HKUST-DXM AI for Finance Joint Laboratory (DXM25EG01). Also supported by National Key Research and Development Program of China Grant No.~2023YFF0725100, National Science Foundation of China (NSFC) under Grant No.~U22B2060, Guangdong-Hong Kong Technology Innovation Joint Funding Scheme Project No.~2024A0505040012, the Hong Kong RGC GRF Project 16213620, RIF Project R6020-19, AOE Project AoE/E-603/18, Theme-based project TRS T41-603/20R, CRF Project C2004-21G, Key Areas Special Project of Guangdong Provincial Universities 2024ZDZX1006, Guangdong Province Science and Technology Plan Project 2023A0505030011, Guangzhou municipality big data intelligence key lab, 2023A03J0012, Hong Kong ITC ITF grants MHX/078/21 and PRP/004/22FX, Zhujiang scholar program 2021JC02X170, Microsoft Research Asia Collaborative Research Grant, HKUST-Webank joint research lab and 2023 HKUST Shenzhen-Hong Kong Collaborative Innovation Institute Green Sustainability Special Fund, from Shui On Xintiandi and the InnoSpace GBA. The authors are grateful to anonymous reviewers for their efforts and insightful suggestions to improve this article.

\bibliography{custom}

\newpage
\appendix
\section{Appendix}
\label{sec:appendix}
\subsection{Sample of Seed Alpha}
\begin{figure}[!htbp]
    \centering
    \includegraphics[width=0.9\linewidth]{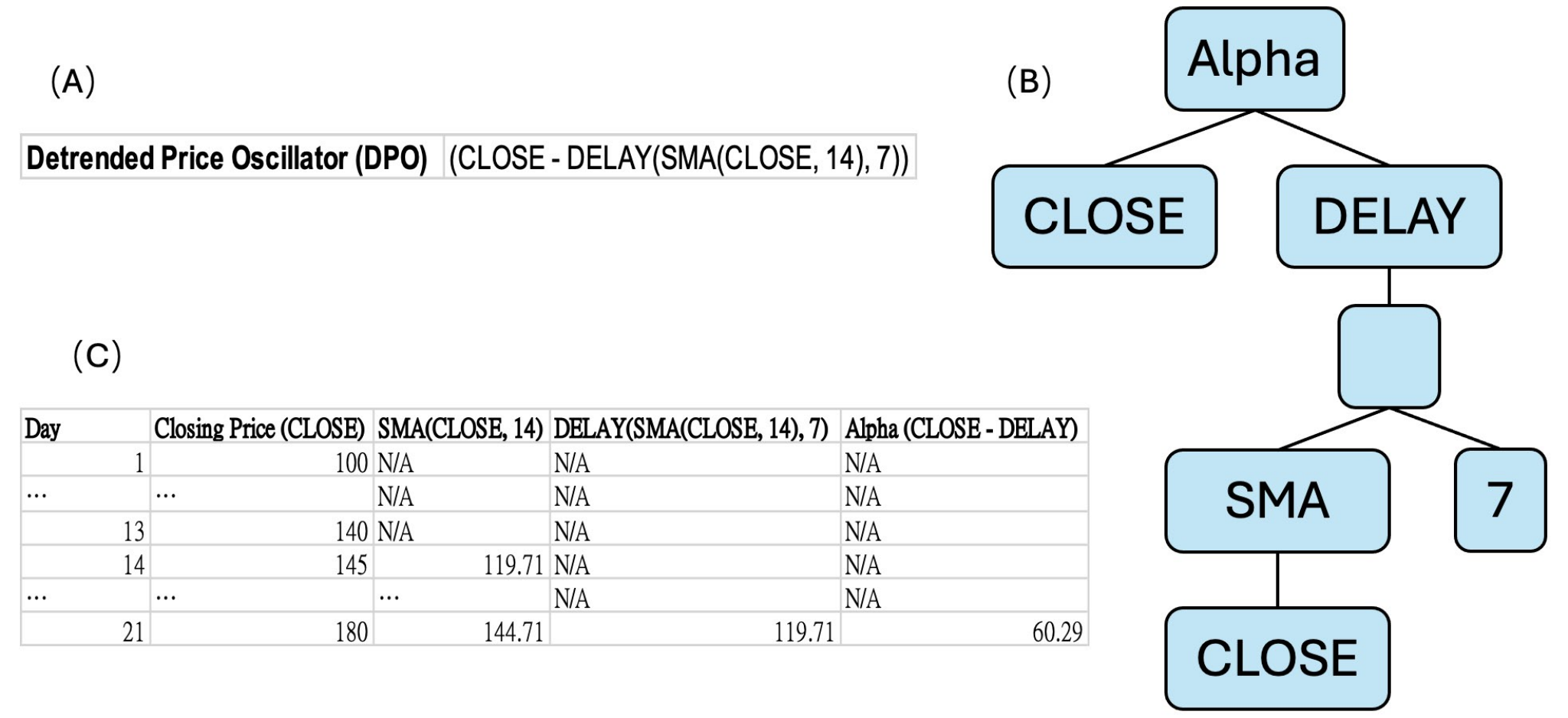}
    \caption{Seed Alpha Representation: (A) An example of the seed alpha formula. (B) Its equivalent expression tree. (C) Step-by-step computation of this seed alpha on an example time series.}
    \label{fig:alpha_sample}
\end{figure}

The seed alpha representation depicted in Figure \ref{fig:alpha_sample} illustrates a fundamental construct in quantitative finance for generating trading signals. This figure presents a comprehensive visualization of the Detrended Price Oscillator (DPO) formula, which compares the current closing price with a delayed simple moving average (SMA) of closing prices. Panel A shows the mathematical formulation of the seed alpha, expressed as \texttt{CLOSE - DELAY(SMA(CLOSE, 14), 7)}, which captures price momentum by measuring deviations from historical trends. Panel B transforms this formula into an equivalent expression tree, demonstrating the hierarchical relationship between operators and operands, which facilitates algorithmic implementation and analysis. Panel C provides a practical illustration through a tabulated step-by-step computation of this alpha formula on a sample time series, showing how the signal evolves over multiple trading days. This representation exemplifies how complex financial indicators can be systematically decomposed, formalized, and applied to market data for quantitative trading strategies.

\begin{table*}[!b]
    \centering
    \resizebox{\linewidth}{!}{
    \begin{tabular}{p{3cm}p{4.5cm}p{9cm}}\hline
Category&Name&Short Code\\ \hline
Momentum&Price Momentum&(CLOSE - DELAY(CLOSE, 14))\\
&Volume Momentum&(VOLUME - DELAY(VOLUME, 14))\\
&RSI Momentum&(RSI - DELAY(RSI, 14))\\
&Rate of Change (ROC)&((CLOSE / DELAY(CLOSE, 14)) - 1)\\
&MACD Momentum&(MACD - DELAY(MACD, 14))\\
&Momentum Oscillator&((CLOSE - DELAY(CLOSE, 14)) / DELAY(CLOSE, 14))\\
&Chande Momentum Oscillator (CMO)&(SUM(IF(CLOSE - DELAY(CLOSE, 1) > 0, CLOSE - DELAY(CLOSE, 1), 0), 14) - SUM(IF(CLOSE - DELAY(CLOSE, 1) $<$ 0, DELAY(CLOSE, 1) - CLOSE, 0), 14)) / (SUM(IF(CLOSE - DELAY(CLOSE, 1) > 0, CLOSE - DELAY(CLOSE, 1), 0), 14) + SUM(IF(CLOSE - DELAY(CLOSE, 1) $<$ 0, DELAY(CLOSE, 1) - CLOSE, 0), 14)) * 100\\
&Stochastic Momentum Index (SMI)&((CLOSE - MIN(LOW, 14)) - (MAX(HIGH, 14) - CLOSE)) / (MAX(HIGH, 14) - MIN(LOW, 14))\\  
&ATR Momentum&(ATR - DELAY(ATR, 14))\\ 
&Detrended Price Oscillator (DPO)&(CLOSE - DELAY(SMA(CLOSE, 14), 7))\\
&Average Directional Index (ADX) Momentum&(ADX - DELAY(ADX, 14))\\ \hline
Mean Reversion&Mean Reversion&(MEAN(CLOSE, 20) - CLOSE)\\
&Z-Score Mean Reversion&(CLOSE - MEAN(CLOSE, 20)) / STD(CLOSE, 20)\\
&Bollinger Bands&(CLOSE - LOWER\_BAND) / (UPPER\_BAND - LOWER\_BAND)\\
&Keltner Channel&(CLOSE - LOWER\_CHANNEL) / (UPPER\_CHANNEL - LOWER\_CHANNEL)\\
&Moving Average Reversion&(SMA(CLOSE, 20) - CLOSE)\\
&Exponential Moving Average (EMA) Reversion&(EMA(CLOSE, 20) - CLOSE)\\
&Distance from High&(MAX(HIGH, 20) - CLOSE)\\
&Distance from Low&(CLOSE - MIN(LOW, 20))\\
&Relative Strength Index (RSI) Reversion&(100 - RSI)\\
&Percent B&((CLOSE - LOWER\_BAND) / (UPPER\_BAND - LOWER\_BAND)) * 100\\
\hline
\end{tabular}}
\end{table*}

\begin{table*}[!t]
    \centering
       \resizebox{\linewidth}{!}{
    \begin{tabular}{p{3cm}p{4.5cm}p{9cm}}\hline
Volatility&Standard Deviation&STD(CLOSE, 20)\\
&Average True Range (ATR)&ATR(14)\\
&Bollinger Band Width&(UPPER\_BAND - LOWER\_BAND) / SMA(CLOSE, 20)\\
&Historical Volatility&STD(RETURNS, 20) * SQRT(252)\\
&Volatility Ratio&STD(CLOSE, 10) / STD(CLOSE, 50)\\
&Chaikin Volatility&(EMA(HIGH - LOW, 10) / DELAY(EMA(HIGH - LOW, 10), 10)) - 1\\
&Garman-Klass Volatility&SQRT(0.5 * LOG(HIGH / LOW)$^2$ - (2 * LOG(2) - 1) * LOG(CLOSE / OPEN)$^2$)\\
&Parkinson Volatility&SQRT((1 / (4 * N * LOG(2))) * SUM(LOG(HIGH / LOW)$^2$, 20))\\
&Yang-Zhang Volatility&SQRT(VAR(LOG(CLOSE / OPEN)) + 0.5 * VAR(LOG(HIGH / OPEN) - LOG(LOW / OPEN)) + 0.25 * VAR(LOG(CLOSE / DELAY(OPEN, 1))))\\
&Ulcer Index&SQRT(MEAN(DRAWDOWN$^2$, 14))\\ \hline
Fundamental&Price-to-Earnings Ratio (P/E)&(CLOSE / EPS)\\
&Price-to-Book Ratio (P/B)&(CLOSE / BOOK\_VALUE)\\
&Dividend Yield&(DIVIDENDS / CLOSE)\\
&Earnings Yield&(EPS / CLOSE)\\
&Sales-to-Price Ratio&(SALES / CLOSE)\\
&Cash Flow Yield&(OPERATING\_CASH\_FLOW / CLOSE)\\ \hline
Liquidity&Trading Volume&VOLUME\\
&Average Trading Volume&MEAN(VOLUME, 20)\\
&Volume Rate of Change (VROC)&(VOLUME - DELAY(VOLUME, 14)) / DELAY(VOLUME, 14)\\
&On-Balance Volume (OBV)&SUM(VOLUME * SIGN(CLOSE - DELAY(CLOSE, 1)))\\
&Liquidity Ratio&VOLUME / MARKET\_CAP\\
&Turnover Rate&VOLUME / SHARES\_OUTSTANDING\\
&Amihud Illiquidity Ratio&ABS(RETURN) / VOLUME\\
&High-Low Spread&(HIGH - LOW) / CLOSE\\
&Dollar Volume&VOLUME * CLOSE\\  
 &Debt-to-Equity Ratio&(TOTAL\_DEBT / TOTAL\_EQUITY)\\
&Return on Equity (ROE)&(NET\_INCOME / EQUITY)\\
&Return on Assets (ROA)&(NET\_INCOME / TOTAL\_ASSETS)\\
&Gross Profit Margin&(GROSS\_PROFIT / REVENUE)\\
&Price-to-Sales Ratio (P/S)&(CLOSE / SALES)\\
&Price-to-Cash Flow Ratio (P/CF)&(CLOSE / OPERATING\_CASH\_FLOW)\\
&Book-to-Market Ratio (B/M)&(BOOK\_VALUE / CLOSE)\\
&Enterprise Value to EBITDA (EV/EBITDA)&(ENTERPRISE\_VALUE / EBITDA)\\
&Bid-Ask Spread&(ASK\_PRICE - BID\_PRICE) / MID\_PRICE\\
&High-Low Spread&(HIGH - LOW) / CLOSE\\
&Dollar Volume&VOLUME * CLOSE\\  \hline
Quality&Gross Profit Margin&(GROSS\_PROFIT / REVENUE)\\
&Operating Profit Margin&(OPERATING\_INCOME / REVENUE)\\
&Net Profit Margin&(NET\_INCOME / REVENUE)\\
&Earnings Stability&STD(EPS, 5) / MEAN(EPS, 5)\\
&Debt to Equity Ratio&(TOTAL\_DEBT / TOTAL\_EQUITY)\\
&Interest Coverage Ratio&(EBIT / INTEREST\_EXPENSE)\\
&Cash Conversion Cycle&(DIO + DSO - DPO)\\
&Asset Turnover Ratio&(REVENUE / TOTAL\_ASSETS)\\ \hline
Growth&Earnings Growth Rate&(EPS / DELAY(EPS, 1) - 1)\\
&Revenue Growth Rate&(REVENUE / DELAY(REVENUE, 1) - 1)\\
&EBITDA Growth Rate&(EBITDA / DELAY(EBITDA, 1) - 1)\\ \hline
\end{tabular}}
\end{table*}

\begin{table*}[!t]
    \centering
       \resizebox{\linewidth}{!}{
    \begin{tabular}{p{3cm}p{4.5cm}p{9cm}}\hline
&Cash Flow Growth Rate&(CASH\_FLOW / DELAY(CASH\_FLOW, 1) - 1)\\
&Dividends Growth Rate&(DIVIDENDS / DELAY(DIVIDENDS, 1) - 1)\\
&Book Value Growth Rate&(BOOK\_VALUE / DELAY(BOOK\_VALUE, 1) - 1)\\
&Sales Growth Rate&(SALES / DELAY(SALES, 1) - 1)\\
&Asset Growth Rate&(ASSETS / DELAY(ASSETS, 1) - 1)\\
&Equity Growth Rate&(EQUITY / DELAY(EQUITY, 1) - 1)\\
&Retained Earnings Growth Rate&(RETAINED\_EARNINGS / DELAY(RETAINED\_EARNINGS, 1) - 1)\\ \hline
Technical&Moving Average (MA)&SMA(CLOSE, 20)\\
&Exponential Moving Average (EMA)&EMA(CLOSE, 20)\\
&Relative Strength Index (RSI)&RSI(14)\\
&Moving Average Convergence Divergence (MACD)&(EMA(CLOSE, 12) - EMA(CLOSE, 26))\\
&Bollinger Bands&UPPER\_BAND - LOWER\_BAND\\
&Stochastic Oscillator&((CLOSE - MIN(LOW, 14)) / (MAX(HIGH, 14) - MIN(LOW, 14))) * 100\\
&Average True Range (ATR)&ATR(14)\\
&Commodity Channel Index (CCI)&(TYPICAL\_PRICE - SMA(TYPICAL\_PRICE, 20)) / (0.015 * MEAN\_DEV(TYPICAL\_PRICE, 20))\\
&Williams \%R&((MAX(HIGH, 14) - CLOSE) / (MAX(HIGH, 14) - MIN(LOW, 14))) * -100\\ \hline
Macro Economics&GDP Growth Rate&GDP - DELAY(GDP, n)\\
&Inflation Rate&CPI - DELAY(CPI, n)\\
&Unemployment Rate&UNEMPLOYMENT\_RATE - DELAY(UNEMPLOYMENT\_RATE, n)\\
&Interest Rate&INTEREST\_RATE - DELAY(INTEREST\_RATE, n)\\
&Industrial Production Index&IPI - DELAY(IPI, n)\\
&Retail Sales Growth&RETAIL\_SALES - DELAY(RETAIL\_SALES, n)\\
&Housing Starts Growth&HOUSING\_STARTS - DELAY(HOUSING\_STARTS, n)\\
&Consumer Confidence Index (CCI)&CCI - DELAY(CCI, n)\\
&Trade Balance&EXPORTS - IMPORTS\\
&Foreign Exchange Reserves&FX\_RESERVES - DELAY(FX\_RESERVES, n)\\
 \hline
\end{tabular}}
\end{table*}
\newpage

\subsection{Seed Alphas document lists for GPTs}
\label{sec:appendix 1}

The presented table offers a comprehensive overview of cutting-edge research pertinent to the development of a SAF in quantitative finance. These documents collectively represent the confluence of traditional financial methodologies with advanced computational techniques. Of particular significance is Kakushadze's "101 Formulaic Alphas," which provides a foundational repository of trading signals that can be categorized according to traditional financial factors. Lopez de Prado's work on "Causal Factor Investing" introduces scientific rigor to factor classification, ensuring independence between alpha categories. The integration of artificial intelligence is evident in multiple studies, including "Alpha-GPT" and "FinGPT," which leverage LLMs for alpha generation. Portfolio construction methodologies are addressed through reinforcement learning approaches in "AlphaPortfolio" and "Dynamic Graph-based Deep Reinforcement Learning." Performance enhancement strategies are explored in "Mastering Stock Markets with Efficient Mixture of Diversified Trading Experts," while market dependency analysis is covered in "Model-Free Implied Dependence and the Cross-Section of Returns." This literature collection provides quantitative researchers with the theoretical frameworks and methodological tools necessary to construct a Seed Alpha Factory that systematically generates, categorizes, and implements trading signals across independent financial categories, balancing traditional financial theory with contemporary computational advancements.

Sample prompt "Summarize the document information to help quantitative researchers build the Seed Alpha Factory according to traditional financial categories, ensuring that each category of seed alphas is independent."

\newpage
\subsection{Generate Seed Alpha factory}
This taxonomy presents quantitative trading signals organized into eight categories: Momentum, Mean Reversion, Volatility, Fundamental, Liquidity, Quality, Growth, Technical, and Macroeconomic indicators. Each category targets distinct market phenomena with specific mathematical implementations. Momentum factors identify persistent trends, Mean Reversion signals detect market overreactions, Volatility metrics quantify price dispersion, and Fundamental factors evaluate company valuations. The framework also includes Liquidity measures of trading activity, Quality indicators of operational efficiency, Growth metrics of financial expansion, Technical indicators derived from price-volume patterns, and Macroeconomic signals reflecting broader economic conditions. The mathematical formulations provided enable researchers to implement diverse, uncorrelated alpha factors for robust quantitative trading strategies.

\label{sec:appendix 2}

\newpage

\subsection{Multimodal Data types}
\label{sec:appendix 3}
Table \ref{Dataset} presents a taxonomy of multimodal data types essential for comprehensive quantitative finance research. The classification encompasses five categories: textual data (financial reports, news articles, social media discourse), numerical data (historical price series, returns, volatility metrics), visual data (charts and technical analysis patterns), audio data (financial broadcasts and market commentary), and video data (specialized financial news channels). This multimodal framework enables researchers to develop more robust predictive models by integrating complementary information channels, potentially identifying market inefficiencies that remain undetectable when analyzing isolated data types. By synthesizing diverse information formats, quantitative analysts can gain multidimensional insights into market dynamics, enhancing both analytical depth and predictive capability in financial modeling applications.
\begin{table}[!htbp]
    \centering
    \caption{Multimodal data types}
    \resizebox{\linewidth}{!}{
    \begin{tabular}{lp{3cm}p{5cm}}\hline
\bf Data Type&\bf Description&\bf Examples\\ \hline
\bf Textual Data&Financial reports, academic papers, news articles, and other textual documents.&Trading forums' sentiment analysis and stock predictions, company disclosures, financial statements, Sina Finance\\
\bf Numerical Data&Historical stock market data, financial metrics, and performance indicators.&Returns, log returns, annualized returns, volatility\\
\bf Visual Data&Charts, graphs, and other visual representations of financial data.&Kline charts, trading charts\\
\bf Audio Data&Financial news broadcasts.&Financial morning news radio, stock review radio, market discussion radio\\
\bf Video Data&Financial news channels.&CCTV Securities Information Channel, CCTV News Broadcast (news affecting China's stock market)
\\ \hline
    \end{tabular}}
    \label{Dataset}
\end{table}

\newpage

\subsection{Category-Based Alpha Selection}
\label{sec:appendix 4}
Algorithm \ref{alg:cap} delineates a systematic methodology for alpha selection in quantitative investment strategies, employing a category-based approach to ensure diversification across multiple financial factors. The procedure operates on a structured set of alpha categories \(\mathcal{C}\), each representing distinct market phenomena such as momentum, mean reversion, or volatility. For each category, the algorithm identifies superior alpha factors through the SelectBestAlphas function, which presumably evaluates historical performance metrics. The innovation lies in the subsequent dual-agent evaluation system: the RiskPreferenceAgent assesses each alpha's risk characteristics, while the ConfidenceScoreAgent evaluates the statistical robustness of its historical performance. These complementary evaluations are synthesized using weight parameters \(w_r\) and \(w_c\) to compute a comprehensive Final Score. Only alphas exceeding a predefined confidence threshold \(X\) are incorporated into the final selection set \(\mathcal{A}\). This methodical approach ensures that the resulting alpha portfolio exhibits both category diversification and individual signal quality, potentially enhancing risk-adjusted returns while mitigating exposure to specific market regimes or factor deterioration.

\begin{algorithm}[!htbp]
\small
\caption{Category-Based Alpha Selection}
\label{alg:cap}
\begin{algorithmic}[1]
\Require Categories \( \mathcal{C} = \{ C_1, \ldots, C_m \} \), each \( C_i \) containing a set of alphas; confidence threshold \( X \)
\State Initialize selected alphas \( \mathcal{A} \gets \emptyset \)
\For{each category \( C_i \in \mathcal{C} \)}
  \State \( \mathcal{A}_i \gets \) SelectBestAlphas(\( C_i \))
  \For{each \( \alpha \in \mathcal{A}_i \)}
    \State \( \text{risk\_score} \gets \) RiskPreferenceAgent(\( \alpha \))
    \State \( \text{confidence\_score} \gets \) ConfidenceScoreAgent(\( \alpha \))
    \State \( \text{final\_score} \gets w_r \cdot \text{risk\_score} + w_c \cdot \text{confidence\_score} \)
    \If{\( \text{final\_score} > X \)}
      \State \( \mathcal{A} \gets \mathcal{A} \cup \{ \alpha \} \)
    \EndIf
  \EndFor
\EndFor
\State \Return \( \mathcal{A} \)
\end{algorithmic}
\end{algorithm}

\newpage

\subsection{Dynamic
Alpha Strategy Construction}
\vspace*{-0.4cm}

\begin{algorithm}[H]
\small
\caption{Overall Framework Algorithm}
\label{alg:overall}
\begin{spacing}{0.97}
\begin{algorithmic}[1]
\Require Multimodal financial data \( \mathcal{D} \), market conditions \( \mathcal{M}^{(t)} \), stocks \( \mathcal{S} \), confidence threshold \( \tau \)
\Ensure Optimized alpha strategy \( \alpha^{(t)} \)

\Statex \textbf{/* Phase 1: SAF */}
\State Initialize empty Seed Alpha Factory \( \mathcal{F} \gets \emptyset \)
\For{each document \( d \in \mathcal{D} \)}
  \State filtered\_content \( \gets \) LLM.Filter(\( d \))
  \State categories \( \gets \) LLM.Categorize(filtered\_content)
  \For{each category \( c \in \) categories}
    \State seed\_alphas \( \gets \) LLM.GenerateAlphas(\( c \))
    \State \( \mathcal{F}_c \gets \mathcal{F}_c \cup \) seed\_alphas
  \EndFor
\EndFor

\Statex \textbf{/* Phase 2: Multi-Agent Evaluation */}
\State Initialize selected alphas \( \mathcal{A} \gets \emptyset \)
\For{each category \( c \in \mathcal{F} \)}
  \For{each alpha \( \alpha_{ij} \in \mathcal{F}_c \)}
    \State \( \theta_{ij} \gets \) ConfidenceScoreAgent(\( \alpha_{ij}, \mathcal{M}^{(t)} \)) \Comment{Statistical evaluation}
    \State \( \rho_{ij} \gets \) RiskPreferenceAgent(\( \alpha_{ij}, \mathcal{M}^{(t)} \)) \Comment{Risk assessment}
    \State \( \text{score}_{ij} \gets w_c \cdot \theta_{ij} + w_r \cdot \rho_{ij} \) \Comment{Combined score}
    \If{\( \text{score}_{ij} > \tau \)}
      \State \( \mathcal{A}_c \gets \mathcal{A}_c \cup \{ \alpha_{ij} \} \)
    \EndIf
  \EndFor
  \If{\( \mathcal{A}_c \neq \emptyset \)}
    \State \( \alpha_c^* \gets \arg\max\limits_{\alpha_{ij} \in \mathcal{A}_c} \text{score}_{ij} \) \Comment{Select best alpha in category}
    \State \( \mathcal{A} \gets \mathcal{A} \cup \{ \alpha_c^* \} \)
  \EndIf
\EndFor

\Statex \textbf{/* Phase 3: Weight Optimization */}
\State Initialize MLP with architecture \( \{ |\mathcal{A}|, 10, 1 \} \) \Comment{Input, hidden, output layers}
\State \( \mathbf{X}_{\text{train}} \gets \) ComputeHistoricalAlphas(\( \mathcal{A} \), \( \mathcal{S} \), \( t_{\text{start}} \), \( t_{\text{end}} \))
\State \( \mathbf{y}_{\text{train}} \gets \) FutureReturns(\( \mathcal{S} \), \( t_{\text{start}} + 1 \), \( t_{\text{end}} + 1 \))
\State \( \mathbf{w} \gets \) TrainMLP(\( \mathbf{X}_{\text{train}} \), \( \mathbf{y}_{\text{train}} \)) \Comment{Learn optimal weights}
\For{each stock \( S_i \in \mathcal{S} \)}
  \State \( \alpha_i^{(t)} \gets \sum_{j=1}^{|\mathcal{A}|} w_j \cdot \alpha_{ij}^{(t)} \) \Comment{Compute composite alpha}
\EndFor
\State \Return \( \alpha^{(t)} = \{ \alpha_1^{(t)}, \alpha_2^{(t)}, \ldots, \alpha_n^{(t)} \} \) \Comment{Final alpha strategy}
\end{algorithmic}
\end{spacing}
\end{algorithm}

\newpage
\subsection{Sample prompt for Seed Alpha selection}
\label{sec:appendix 5}

\quad 1. The training set includes:
\begin{itemize}
    \item \textbf{Financial Reports:} 6 quarters (from Jan 2021 to Jun 2022) for 50 companies listed in the SSE 50.
\item  \textbf{Factor Analysis Data:} 37 factors (from Jan 2021, to Jun 2022) divided into groups: Momentum, Mean Reversion, Volatility, Fundamental, Quality, Growth, Technical, Macro Economics. The metric used is the IC.
\end{itemize}

2. \textbf{Objective:} Learn the relationship between the performance of financial reports for the first four quarters from 2022 to 2023 and the factor analysis data (IC) for each of the 37 factors in the last quarter of 2022.

3. \textbf{When provided with the test set (performance of the first 4 quarters of 2023):}
\begin{itemize}
    \item Select the factors that will perform best in the last quarter of the SSE 50.
\item Provide a confidence score \& the risk preference for your selection for each selected Alpha factor.
\end{itemize}

4. \textbf{Selection Criteria:}
\begin{itemize}
    \item If no relationship between financial reports and IC can be found, select the Alpha factor with the highest IC value in each group.
\item For verification of market information differences, if no relationship between financial reports and IR can be found, select the Alpha factor with the highest IR value.
\end{itemize}

\newpage
\subsection{Portfolio Construction Methods}
\label{portfolio construction}
Our investment methodology implements a daily portfolio reconstruction using a top-$k$/drop-$n$ selection framework. At each trading session, we rank all securities by their alpha values—quantitative indicators of expected excess returns—and select the top $k$ stocks for portfolio inclusion. This approach targets securities with the strongest signals, potentially exploiting short-term market inefficiencies. We employ an equal-weighting scheme across selected securities, distributing capital homogeneously among the top candidates, which offers diversification benefits and aligns with our view that alpha signals primarily provide directional indications rather than precise return forecasts.

To optimize transaction costs and operational efficiency, we limit portfolio turnover to a maximum of $n$ securities per trading day. This constraint balances maintaining alignment with current alpha signals while minimizing trading friction costs. For our experiment in Section~\ref{4.4}, we set $k = 13$ and $n = 5$ based on extensive backtesting. This configuration establishes a portfolio concentration that balances diversification against signal dilution, while limiting daily turnover to approximately 38\%—a level that demonstrates favorable characteristics in our transaction cost modeling and represents an efficient trade-off between signal utilization and implementation costs.

\end{document}